\documentclass[english,a4paper,14pt]{article}
\usepackage[
  a4paper, mag=1000, includefoot,
  left=3cm, right=1cm, top=2cm, bottom=2cm, headsep=1cm, footskip=1cm
]{geometry}
\usepackage[T2A]{fontenc}
\usepackage[utf8]{inputenc}

\usepackage[pdfencoding=auto]{hyperref}
\usepackage{xcolor}
\hypersetup{
    colorlinks,
    linkcolor={red!50!black},
    citecolor={blue!50!black},
    urlcolor={blue!80!black}
}
\usepackage{amsmath}
\usepackage{gensymb}
\usepackage{amssymb}
\usepackage{relsize}
\usepackage{array}
\usepackage{lipsum}
\usepackage{tabularx}
\usepackage{graphicx}
\usepackage{cite}
\usepackage{subcaption}
\usepackage{multirow}
\usepackage{booktabs}
\usepackage{empheq}

\title{Solute diffusion calculation in Fe-Si and Fe-Cr-Si multicomponent alloys}
\date{\vspace{-3ex}}

\usepackage{indentfirst}

\begin{document}
\maketitle


\vspace{-40pt}
\section*{Authors}
\noindent $\text{{\bf Miryashkin Т. N.}}^{1*}$, $\text{{Novoselov I. I.}}^{2}$, $\text{{Yanilkin A. V.}}^{2,1}$ 
\\ \\
$^1$ - Moscow Institute of Physics and Technology, Institutskiy Pereulok 9, Dolgoprudny, Moscow Region 141700, Russia
\\
$^2$ - Dukhov Automatics Research Institute (VNIIA), Sushchevskaya 22, Moscow 127055, Russia
\\
*t.miryashkin@gmail.com; \\

\vspace{-20pt}
\section*{Keywords}
\noindent Solute diffusion, bond potential, low-rank potential, correlation factor.

\vspace{-10pt}
\section*{Abstract}

Diffusion plays a key role in microstructure evolution at multicomponent alloys: diffusion controls the kinetics of phase transformations and alloy homogenization. This study aims at developing computationally efficient approaches to estimate the solute diffusion coefficients in two-component systems. We consider silicon as the solute example because it is highly used in industrial steels. We demonstrate that the silicon jump frequency may be calculated with the bond potential instead of the more computationally expensive machine learning potential in Fe-Si and Fe-Cr-Si alloys. We show that the silicon jump frequency can be estimated from thermodynamic simulations for the bond potential without kinetic simulations. The silicon correlation factor slightly depends on silicon concentration and can be approximately estimated by the analytical nine-frequency model.

\section{Introduction}

Industry extensively uses steel as a construction material. Steel consists of an iron matrix and various alloying elements; alloying elements give specific properties to steel: hardness, durability, corrosion resistance. Diffusion of the alloying elements plays a crucial role in the kinetics of many microstructural changes: new phases nucleation, phase transformations, alloys homogenization \cite{Fultz_2014, Calphad_review, Kelton_book}. The experimental studies of solute diffusion in materials are resource-intensive and time-consuming. The development of computational methods and analytical models that estimate diffusion helps to efficiently design steel with specific properties for various industrial needs.

The analytical models can be used to estimate the solute diffusion in infinitely diluted binary alloys \cite{Manning1964, LeClaire_Jones_1972, mehrer_2007}. These models require binding energy and migration energy, which can be estimated from the ab-initio simulations \cite{Vincent_DFT_BindEn, Gorbatov_DFT_BindEn, Olsson_DFT_BindEn, Messina_2014} or from the experiments \cite{Experiment_DFT_BindEn, vac_mig_ener_exper}. In real work steels, the typical concentrations of the alloying element reach up to 1 - 1.5 at. \% \cite{steels_nucler_reactors}; thus, we are required to account for the corresponding effects that interactions of the solute atoms produce \cite{mehrer_2007}. The Darken and Manning theories describe how the solute diffusion changes when the solute concentration is increased \cite{Darken_theory, Manning_theory} and account for the contributions from thermodynamic and kinetic factors. These theories were thoroughly analyzed and compared to atomic kinetic Monte Carlo simulations with pairwise interaction potential in \cite{Swoboda2010}. The main drawback is that these theories do not consider changes in the hopping barriers due to the increasing concentration of solute atoms, which lead to exponential changes in diffusion constants and cannot be neglected in the full analysis.

The computational approaches estimate the diffusion coefficients more accurately than the approximate analytical models. The major atomistic modeling methods that are used to study diffusion include the ab initio molecular dynamics, classical molecular dynamics, Monte Carlo methods. Ab initio molecular dynamics accurately reproduces energy and forces in the system but is computationally expensive \cite{ComputationalQC_Cancs2003, Finnis_interatomicForces}. Classical molecular dynamics requires interatomic potentials. Semiempirical potentials cannot provide the desired accuracy for describing multicomponent systems, while machine learning interatomic potentials need a tremendous amount of configurations for training \cite{Shapeev_mtp, GAP_2010, Behler_NN_Potential, MLP_Cost}. Moreover, the total simulation time for classical molecular dynamics is typically limited to several microseconds as the timesteps should be of $ \sim 10^{-15}$ s to accurately integrate the equations of motion. Interatomic potentials for the Monte Carlo simulations are constructed using sufficiently fewer configurations \cite{Shapeev_LRP_2017}. The kinetic Monte Carlo method may be used for several orders longer simulations compared to the molecular dynamics \cite{Voter_KMC, Chatterjee2007}.

This paper aims to develop computationally efficient methods to estimate solute diffusion in two-component alloys on the example of the Fe-Si system. Firstly, the current work investigates the application of the bond potential instead of the more expensive machine learning potential. Secondly, we explore how to estimate the solute diffusion coefficient from thermodynamic modeling without requiring additional kinetic simulations. We use the kinetic Monte Carlo method to conduct kinetic simulations and to sample equilibrium configuration for the thermodynamics computations.

We investigate Fe-Si (and Fe-Cr-Si) system at 800 K, where the iron is ferromagnetic; the Curie temperature for iron is 1043 К \cite{kittel_solid_state_physics_8}. We select Si as an example of alloying element because industry uses Si to increase corrosion resistance and to strengthen the steel. The selected temperature is particularly important as steels in nuclear reactors operate under this temperature \cite{reactor_struct_materials}. Understanding the physics at this temperature is crucial to guarantee the safety of power plants. We are unaware of experiments for silicon diffusion in Fe-Si (and Fe-Cr-Si) system for the ferromagnetic phase of Fe. The experimental works we found investigate the higher temperatures, where the iron is paramagnetic \cite{feSi_diffusion_experiments, feCr_diff_exper1, feCr_diff_exper2}. We do not extrapolate the results of these experiments to lower temperatures, as the phase transition of Fe arguably changes the physics of the diffusion. Future experiments and computational works for lower temperatures should clarify the physics of ferromagnetic iron.

We organize the article in the following way. We describe theoretical methods and computational details in Section \ref{sec:comp_details}. We present the results for the silicon jump frequency and correlation factor in Fe-Si alloy in Section \ref{sec:Results}. We summarize our results in Section \ref{sec:conclusions}. In the Appendix, we investigate silicon jump frequency and correlation factor in Fe-Cr-Si alloy with Cr concentration up to 12 at.\%.


\section{Computational Details}
\label{sec:comp_details}

\subsection{Solute Diffusion}
\label{subsec:sol_dif_coef}

Solute diffusion coefficient ($D_{\rm sol}$) in cubic lattices is expressed by the formula: 

\begin{equation}
    D (C_{\rm sol}) = \frac{\Delta^2}{6} f({C_{\rm sol}}) \, \nu ({C_{\rm  sol}}) \, C_V ({C_{\rm sol}}), 
    \label{eq:D2_mainEq}        
\end{equation}
where $\Delta$ is the distance between the nearest neighbors, $f$ is the correlation factor, $\nu$ is the solute jump frequency, $C_V$ is the vacancy concentration, $C_{\rm sol}$ is the solute concentration. We consider the vacancy diffusion mechanism as existing research shows it is a  dominant mechanism for steels \cite{selfDiffMech_overview, mehrer_2007}. 

The present article considers the solute jump frequency and correlation factor. We describe how one can calculate the solute jump frequency from the thermodynamic simulations without requiring direct kinetic simulations for the bond potential (Section \ref{sec:thermo_nu_si}). In the Results (Section \ref{sec:Results}), we estimate the correlation factor from the analytical models \cite{Le_Claire_1970, LeClaire_Jones_1972, Manning1964} and show how it depends on solute concentration. Calculation of vacancy concentration from thermodynamic Monte Carlo simulations is established routine and can be found in the articles \cite{point_defect_thermodyn, thermodyn_hydrogen_vac}.

\subsection{Interatomic Potentials}
\label{sec:iterat_pot}

We select the low-rank potential \cite{Shapeev_LRP_2017, Kostiuchenko_lrp} and bond potential \cite{kmc_mig_barrier_2} as the interaction models in our computations. Kinetic Monte Carlo simulations with the low-rank potential serve as reference simulations, as the authors in \cite{MeshkovNovoselov2020} demonstrate that the low-rank potential is reliable for diffusion calculations. We propose the bond potential instead of the low-rank potential, as the bond potential is more computationally efficient and may be easily constructed on the several quantum-mechanical computations.

The low-rank potential partitions the energy into contributions of the atomic environments and represents these contributions with low-rank multidimensional tensors \cite{tt_oseledets}. This way the low-rank potential accounts for multiparticle interactions and captures the atomic interactions up to the fifth coordination sphere. The potential parameters are trained by minimizing the root-mean-square error. The training dataset consist of the configurations that we calculate in the density functional theory framework with projector augmented wave method \cite{PAW_potentials} using Vienna Ab-Initio Simulation Package (VASP) \cite{VASP_1, VASP_2}. 

\begin{figure}[h!]
\begin{center}
	\includegraphics[width=0.99\textwidth]{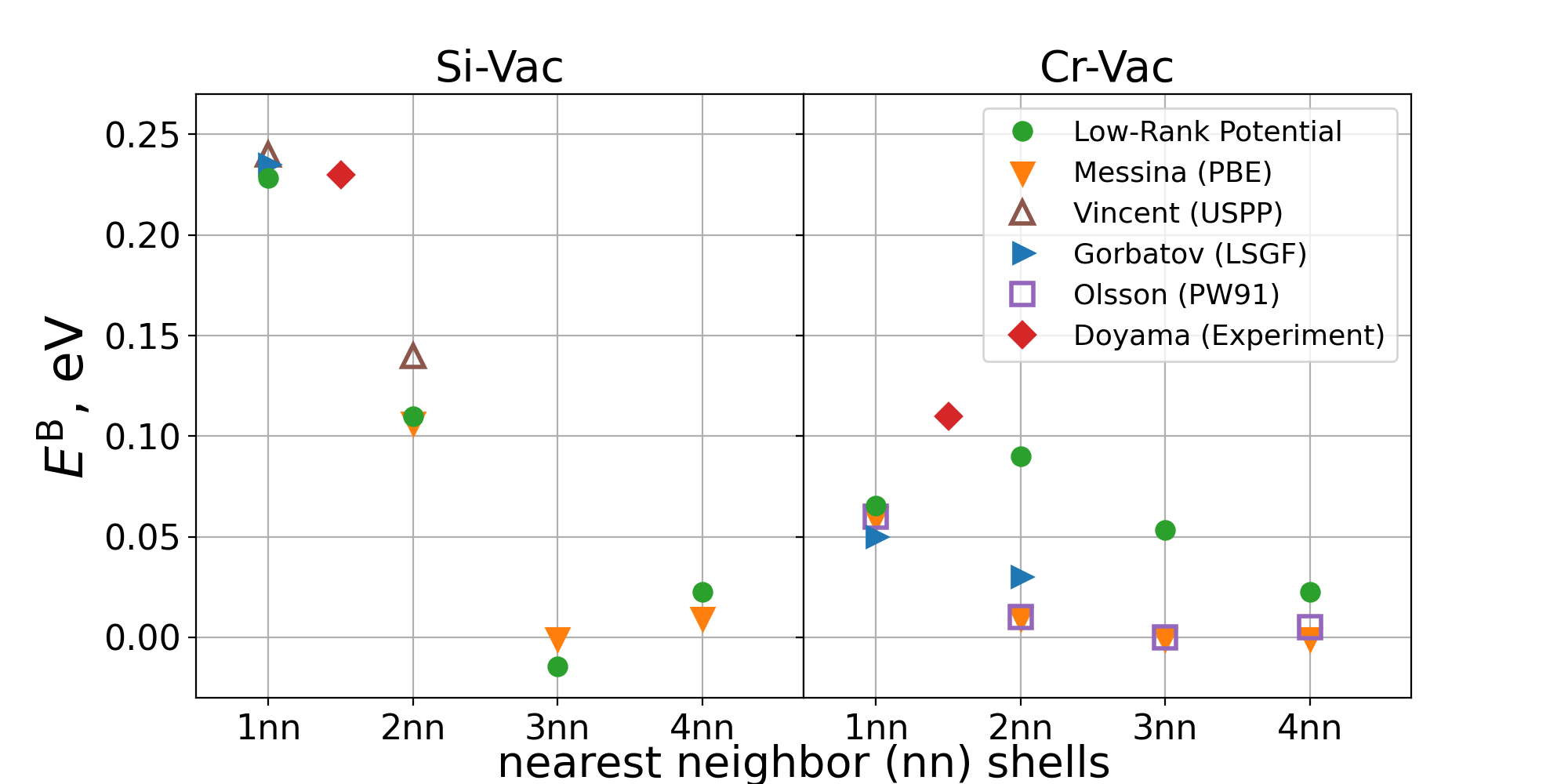}	
	\caption{We compare binding energies from the low-rank potential to the literature values. Notations in the figure indicate the following works: Messina \cite{Messina_2014}, Vincent \cite{Vincent_DFT_BindEn}, Gorbatov \cite{Gorbatov_DFT_BindEn}, Olsson  \cite{Olsson_DFT_BindEn}, Doyama \cite{Experiment_DFT_BindEn}.}
	\label{fig:Eb_LRP_Ref_Comparas}
\end{center}
\end{figure}

Using the low-rank potential, we calculate the binding energies, which are presented in Table \ref{tab:energiesBondModel}. We compare silicon-vacancy (Si-Vac) and chromium-vacancy (Cr-Vac) binding energies with the density functional theory calculations and experimental data in Figure \ref{fig:Eb_LRP_Ref_Comparas}. The obtained results for the Si-Vac binding energies are in good agreement with the previous works. However, the Cr-Vac binding energies in 2-nd and 3-rd coordination shells differ from the available data. The low-rank potential is trained on structures with Cr concentrations up to 14 atomic \% (at.\%) and Si up to 2 at.\%. The Cr-Vac binding energies discrepancy might occur as the low-rank potential is designed to describe a broader range of Cr concentrations and hence works worse for the case of low concentration.

\begin{table}[h!]
\begin{center}
\begin{tabular}{crrrr}
\toprule
{} &     1 NN &     2 NN &     3 NN &     4 NN \\
\midrule
Fe - X   &  \multirow{2}{*}{0.000} &  \multirow{2}{*}{0.000} &  \multirow{2}{*}{0.000} &  \multirow{2}{*}{0.000} \\ 
(X = Fe, Cr, Si, Vac)   &  &  &  &  \\
Cr - Cr   &  -0.105 &  -0.082 &  -0.025 &  -0.027 \\
Si - Cr   &  -0.089 &  -0.016 &  -0.049 &  -0.017 \\
Si - Si   &  -0.390 &  -0.150 & 0.002 & 0.009 \\
Cr - Vac  & 0.066 & 0.090 & 0.053 & 0.023 \\
Si - Vac  & 0.228 & 0.110 &  -0.014 & 0.023 \\
\bottomrule
\end{tabular}
\caption{Binding energies calculated from the low-rank potential. 'NN' denotes nearest neighbor shell.}
\label{tab:energiesBondModel}
\end{center}
\end{table}

The bond potential accounts for the pair-wise interactions between nearest-neighbor lattice sites, which point defects or atoms may occupy \cite{kmc_mig_barrier_2}. The main advantage of the bond potential is that it can be easily constructed for alloys with different elements on tens of quantum-mechanical calculations \cite{TheBondModel_IronStudy}. At the same time, the low-rank potential requires hundreds of quantum-mechanical calculations for training \cite{Shapeev_LRP_2017}. Moreover, the bond potential is more efficient during the simulations. For the bond potential, the system energy is updated by accounting for the bonds that change after the diffusive jump. In the low-rank potential implementation, contributions of the atomic environments must be updated up to the fifth coordination shell, which is more computationally demanding.

We parameterize the bond potential on the binding energies taken from the first column of Table \ref{tab:energiesBondModel}, as we intend to compare results obtained with the bond and low-rank potentials. Although, one may take the binding energies directly from quantum-mechanical calculations.

\subsection{Improved Equation for Solute Jump Frequency}
\label{sec:thermo_nu_si}

In the current section, we describe how one can calculate the solute jump frequency from the thermodynamic simulations without requiring direct kinetic simulations. As the first approximation, we use the following formula to estimate the solute jump frequency from the thermodynamic modeling:  

\begin{equation}
    \nu_{\rm Sol} (C_{\rm Sol}) = \omega_0 \exp \left( - \beta E_{\rm mig} \right) \,  \exp \left( - \beta E_{\rm Sol-V} (C_{\rm Sol}) \right), 
    \label{eq:nuSol_inf_diluted}
\end{equation}
where $\omega_0$ is the attempt jump frequency,  $E_{\rm mig}$ is the migration energy of solute, $E_{\rm Sol-V} (C_{\rm Sol})$ is the solute-vacancy binding energy, $\beta$ is the inverse temperature. This equation reduce to the Lomer equation \cite{mehrer_2007_FFM} for the case $C_{\rm Sol} \rightarrow 0$.

We improve the Equation \eqref{eq:nuSol_inf_diluted} to more accurately estimate solute jump frequency ($ \nu_{\rm Si} $) for the bond potential. We consider adjusting the migration energy, which accounts for local environment properties. The system's energies before and after the diffusive jump are denoted as $E_{\rm I}$ (initial energy) and $E_{\rm F}$ (final energy). So the adjustment for migration energy has the form:

\begin{equation}
  \Delta E_{\rm mig} = \frac{E_{\rm F}(C_{\rm Sol}) - E_{\rm I}(C_{\rm Sol})}{2}. 
  \label{eq:delta_e_mig}
\end{equation}

For the case of the bond potential, $\Delta E_{\rm mig}$ may be further simplified by considering the number of solute-vacancy and solute-solute bonds before and after the jump: 

\begin{equation}
  \Delta E_{\rm mig} = \frac{1}{2} (N_{\rm Sol} - 1) \left(  \, E_{\rm Sol-Sol} - \, E_{\rm Sol-V} \right).   
  \label{eq:delta_e_mig_bond_pot} 
\end{equation}
 
In the present work, we consider Fe-Si alloy with the body-centered cubic lattice. For the body-centered cubic lattice, there are eight possible vacancy energy configurations, which differ by the number of solute atoms in the first coordination shell ($N_{\rm Sol} \in \{1, 2, \ldots, 8\} $). We define the probability of finding the particular energy configuration by the $ p (N_{\rm Sol}) $ coefficients. These probabilities can be obtained from the thermodynamic modeling by counting the occurrences of each energy configuration. The improved formula for the solute jump frequency is given by 

\begin{equation}
  \nu_{\rm Sol} = \omega_0 \exp \left( - \beta E_{\rm mig} \right) \sum_{N_{\rm Sol} \in \{1, 2, \ldots, 8\}} p(N_{\rm Sol}) \exp \left[- \beta \left( E_{\rm Sol-V} (N_{\rm Sol}) - \Delta E_{\rm mig} (N_{\rm Sol}) \right) \right].
  \label{eq:nu_si_bond}
\end{equation}

\subsection{Solute-Vacancy Binding Energy}

We estimate the solute-vacancy binding energy ($E_{\rm Sol-V} (C_{\rm Sol})$) from thermodynamic modeling by the following formula: 

\begin{equation}
  E_{\rm Sol-V} (C_{\rm Sol}) = \mu_{\rm Sol-V} (C_{\rm Sol}) - \mu_{V} (C_{\rm Sol}) ,
  \label{eq:eff_bind_energy}
\end{equation}
where $\mu_{V}$ is the vacancy's bulk chemical potential, $\mu_{\rm Sol-V}$ is the vacancy's chemical potential with one or more solute atoms in the first coordination shell.

The procedure for calculating chemical potentials in Equation \eqref{eq:eff_bind_energy} is as follows. Firstly, we estimate the chemical potentials of Fe ($\mu_{\rm Fe}$) and Sol ($\mu_{\rm Sol}$) in Fe-Sol system:

\begin{empheq}[left=\empheqlbrace]{alignat=2}
  E &= \mu_{\rm Fe} N_{\rm Fe} + \mu_{\rm Sol} N_{\rm Sol}, \\ 
      \Delta^{\rm Fe \rightarrow  Sol} E &= \mu_{\rm Sol} - \mu_{\rm Fe},
\end{empheq}
where $E$ is the total system's energy, $ \Delta^{\rm Fe \rightarrow Sol} E $ is the substitution energy of Fe to Sol atom, $N_{\rm Fe}$ and $N_{\rm Sol}$ are the numbers of the Fe and Sol atoms. Then, we calculate a vacancy's chemical potentials:

\begin{equation}
\mu_{V} = \mu_{\rm Fe} - \Delta^{\rm V \rightarrow Fe} E,   
\label{eq:mu_V_sub_ener}
\end{equation}

\begin{equation}
\mu_{\rm Sol-V} = \mu_{\rm Fe} - \Delta^{\rm V \rightarrow Fe}_{\rm Sol} E,  
\label{eq:mu_V-Si_sub_ener}
\end{equation}
where $ \Delta^{\rm V \rightarrow Fe} E $ is the substitution energy of vacancy to Fe site, $\Delta^{\rm V \rightarrow Fe}_{\rm Sol} E$ is the  substitution energy of vacancy to Fe site, which has one or more Sol atoms in the first coordination shell. 

Substitutional energies are averaged according to the formula:

\begin{equation}
  \langle \Delta E \rangle  = -k_{\rm B} T \log \left( \left \langle \exp \left( - \frac{\Delta E}{k_{\rm B} T} \right) \right \rangle  \right) .
  \label{eq:deltaE_Averaging}
\end{equation}

\subsection{Simulation Details}

We use atomic kinetic Monte Carlo method to conduct direct kinetic simulations and to sample equilibrium configuration for the thermodynamics computations. The Stochastic Parallel Particle Kinetic Simulator (SPPARKS) \cite{Spparks_2008} is used for simulations. The sequence of system configurations is generated using the algorithm by Metropolis \cite{MetropolisAlgorithm_1953}. 

The migration energies are adopted from \cite{Messina_2014}: $E_{\rm mig}^{\rm Fe} = 0.70 $ eV, $E_{\rm mig}^{\rm Si} = 0.51$ eV, $E_{\rm mig}^{\rm Cr} = 0.53 $ eV. These energies are obtained using the density function theory framework with the generalized gradient approximation \cite{GGA_pseudopot}. The attempt frequency ($\omega_0$) is selected of the order of the Debye frequency in Fe, which is 6 THz, and is assumed to be the same for all types of atoms.


\section{Results}
\label{sec:Results} 

\subsection{Solute Jump Frequency}
\label{subsec:nuSi_feSi}

Here we apply formulas from Section \ref{sec:thermo_nu_si} to estimate the solute jump frequency from the thermodynamic simulations and compare the results with the direct atomic kinetic Monte Carlo simulations. 

\begin{figure}[h!]
\begin{center}
	\includegraphics[width=0.7\textwidth]{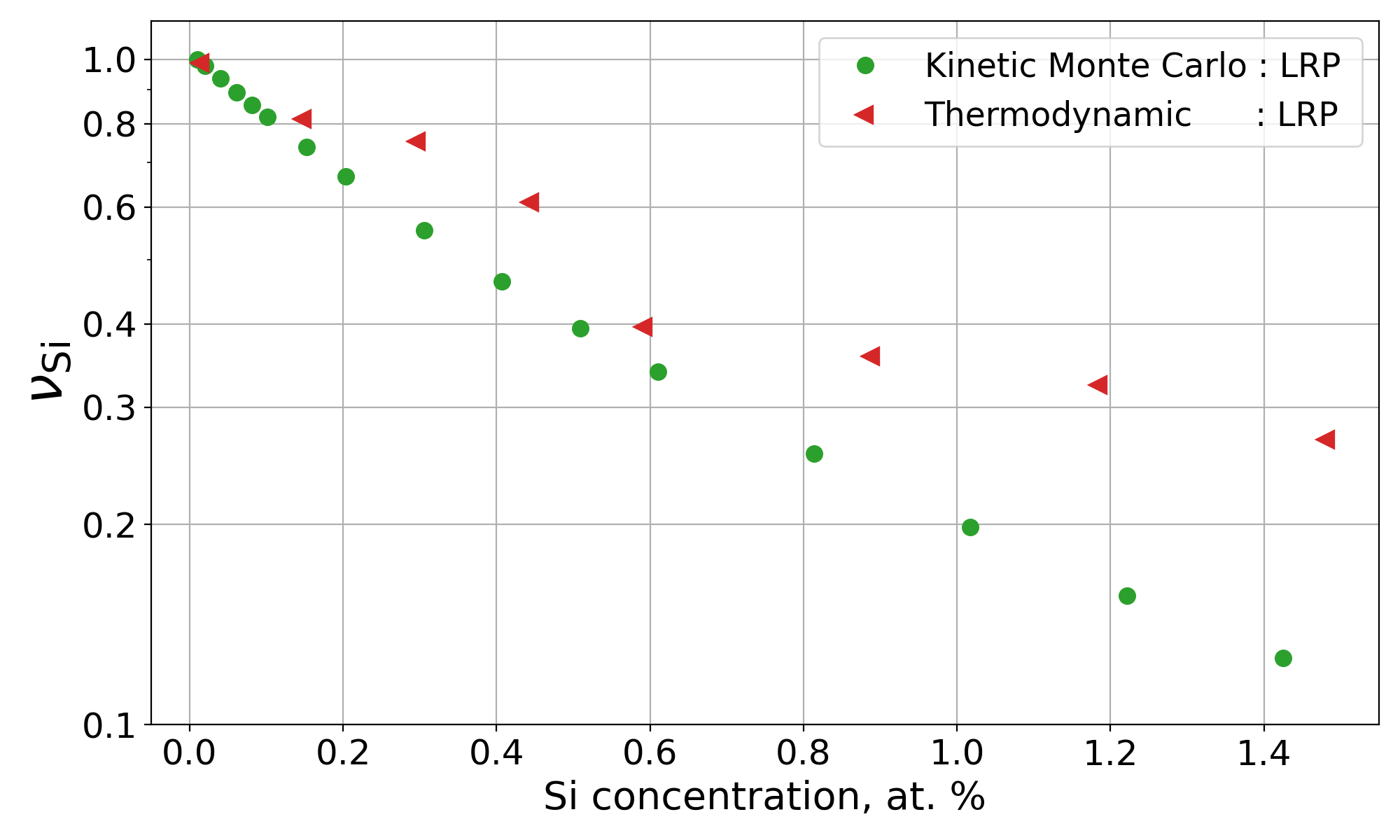}	
	\caption{Si jump frequency ($\nu_{\rm Si}$) dependence on Si concentration in Fe-Si system. We take $\nu_{\rm Si}$ from direct kinetic simulations and thermodynamic estimations of Si-Vac binding energy ($E_{\rm Si-V} (C_{\rm Si})$). Both methods use the low-rank potential (LRP).}
	\label{fig:nuSi_SimWidom_LRP}
\end{center}
\end{figure}

We start with the simulations performed using the low-rank potential. In Figure \ref{fig:nuSi_SimWidom_LRP}, we compare two approaches: direct atomic kinetic Monte Carlo simulations and thermodynamic method utilizing solute-vacancy binding energy $(E_{\rm Si-V} (C_{\rm Si}))$. We norm silicon jump frequency ($\nu_{\rm Si}$) to the frequency in the infinitely diluted system. Thermodynamic calculations by Equation \eqref{eq:nuSol_inf_diluted} capture the dependence of $\nu_{\rm Si}$; although the thermodynamic modeling provides higher values of $\nu_{\rm Si}$ compared to the reference kinetic simulations. The Equation \eqref{eq:nuSol_inf_diluted} can be used to qualitatively estimate the jump frequency for the low-rank potential but requires further improvements to reproduce the kinetic results.

\begin{figure}
    \centering
    \begin{subfigure}[b]{0.65\linewidth}
        \includegraphics[width=\linewidth]{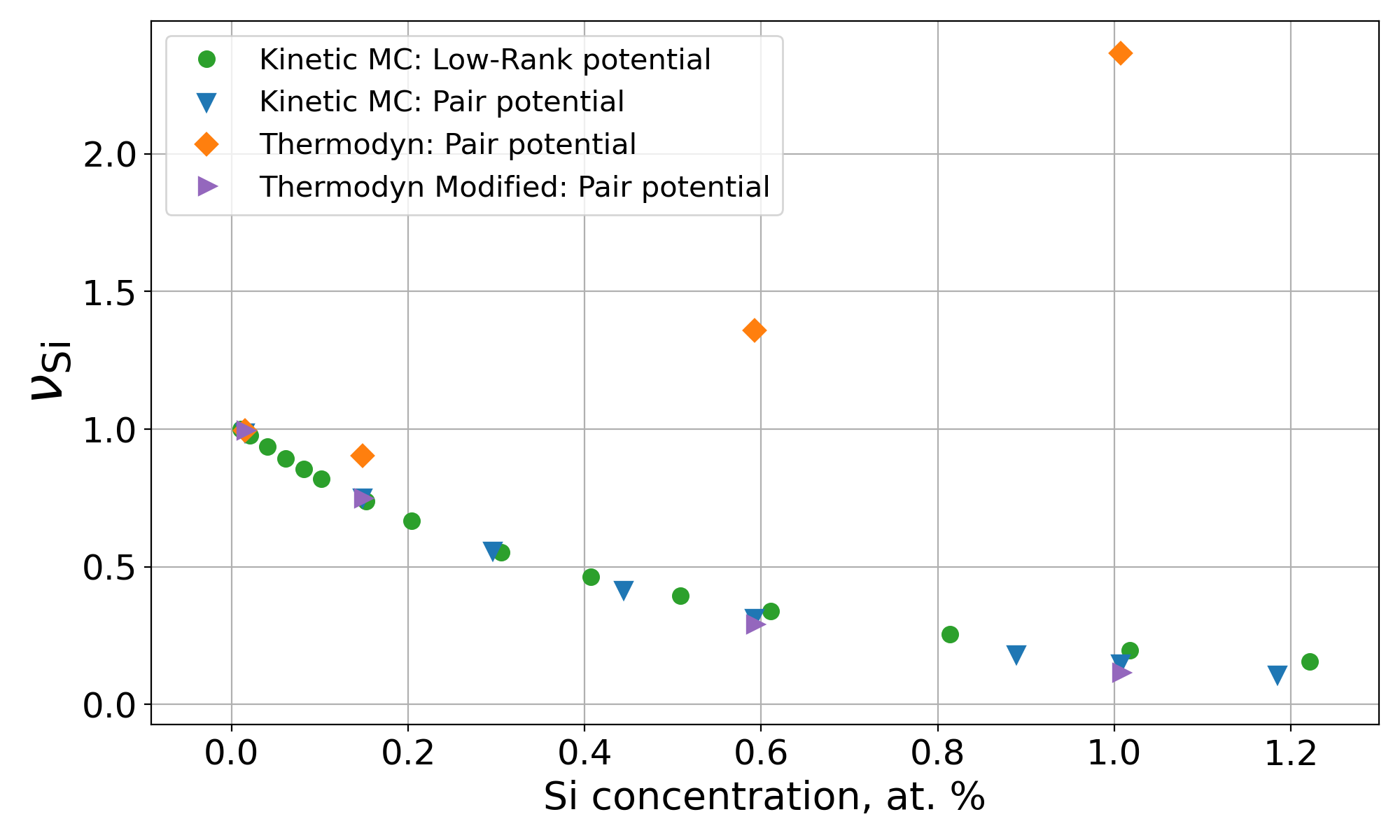}
        \caption{}
        \label{subfig:nuSi_bondModel_1}
    \end{subfigure} \\
    \begin{subfigure}[b]{0.65\linewidth}
        \includegraphics[width=\linewidth]{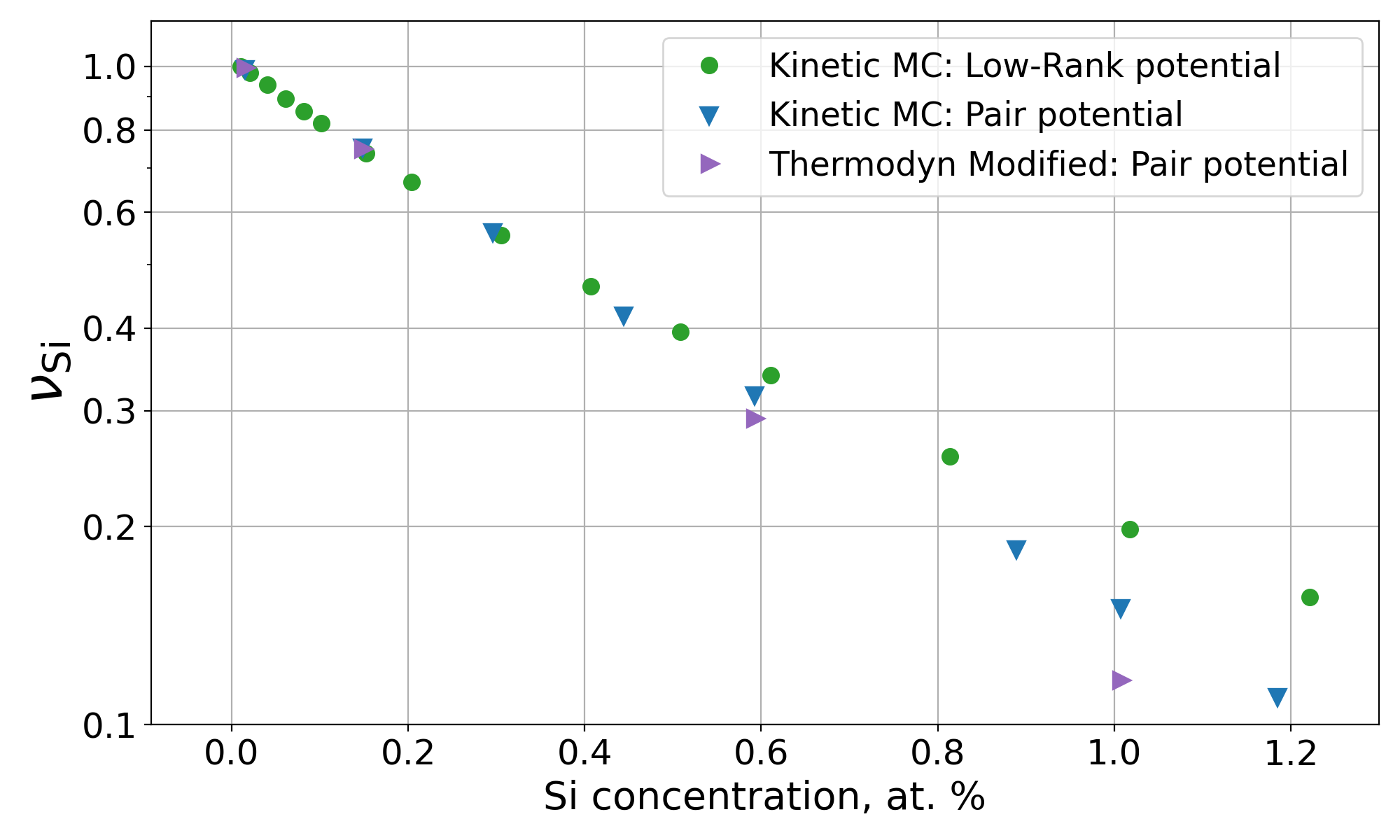}       
        \caption{}
        \label{subfig:nuSi_bondModel_2}
    \end{subfigure}
	\caption{Silicon jump frequency estimated by kinetic and thermodynamic approaches. Thermodynamic calculations by the approximate Equation \eqref{eq:nuSol_inf_diluted} are shown in Figure \ref{subfig:nuSi_bondModel_1}. Figure \ref{subfig:nuSi_bondModel_2} presents the results obtained by the improved Equation \eqref{eq:nu_si_bond}.}
    \label{fig:nuSi_bondModel} 
\end{figure}

We further use the bond potential to compare the calculations by both methods: thermodynamic and kinetic ones. Figure \ref{subfig:nuSi_bondModel_1} presents two significant results. Firstly, the direct kinetic simulations with the bond potential are in good agreement with the low-rank potential simulations. This agreement indicates that the diffusion computations can be sufficiently speeded up using the bond potential. The possibility for such acceleration is significant for industrial applications, so we carried out the simulations in stainless steel, which is an essential constructional material (see Appendix). Secondly, the approximate Equation \eqref{eq:nuSol_inf_diluted} does not work for the bond potential and produces physically inadequate results. 

\begin{figure}[h!]
    \begin{center}
        \includegraphics[width=0.6\textwidth]{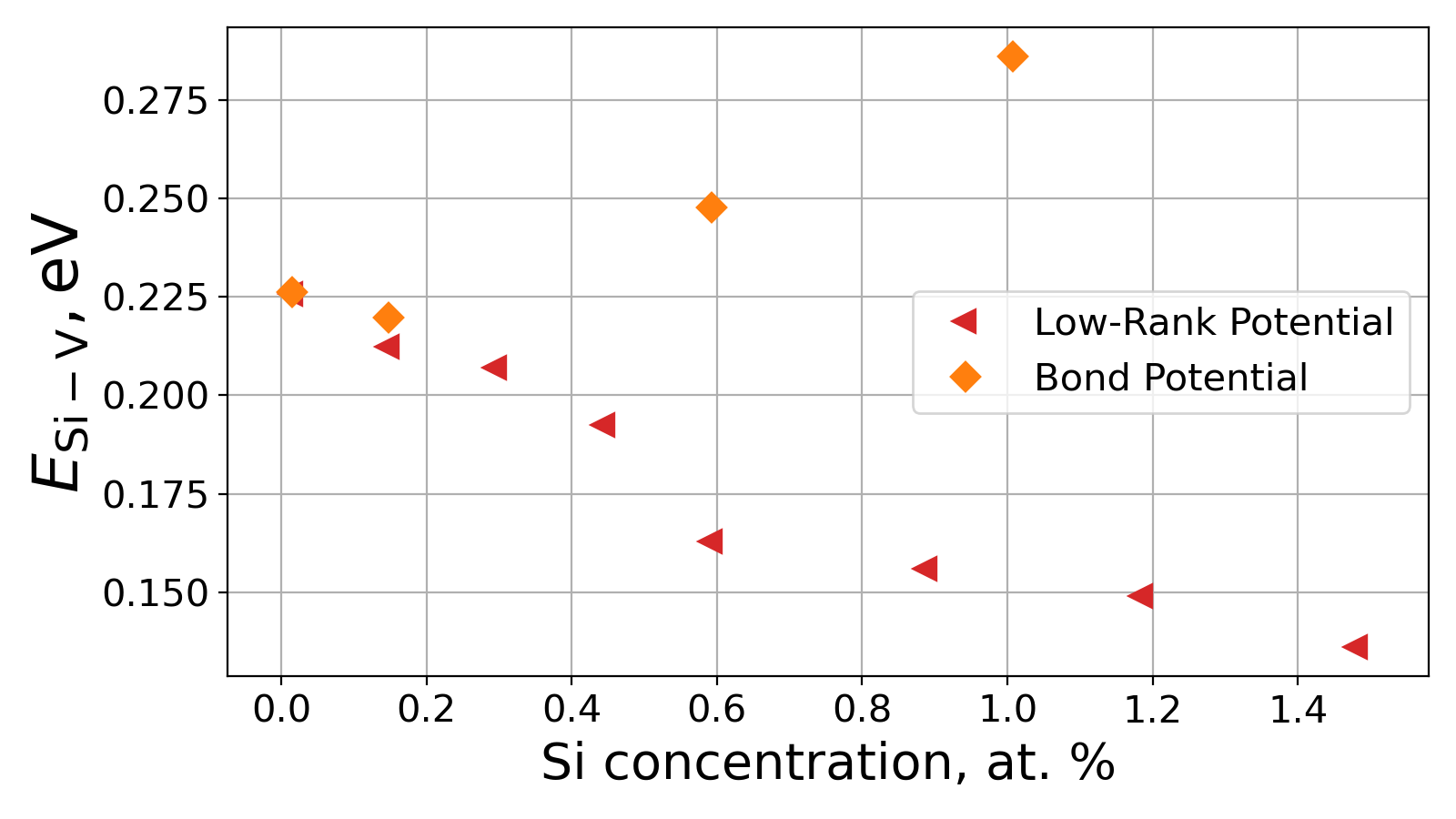}	
        \caption{Silicon-vacancy binding energy ($ E_{\rm Si-V} (C_{\rm Si})$)  calculated using the low-rank potential and the bond potential.}
        \label{fig:dmu_ConcDep}
    \end{center}
\end{figure} 

We analyze the Si-Vac binding energies in Figure \ref{fig:dmu_ConcDep} to understand these inadequate results in order to further improve Equation \eqref{eq:nuSol_inf_diluted}. For the low-rank potential, $E_{\rm Si-V}$ gradually decreases with the increment of silicon concentration. When the silicon concentration is increased, more silicon atoms occupy the sites in the vacancy's first coordination shell. Table \ref{tab:energiesBondModel} shows that silicon atoms repel in the second coordination shell; thus, we observe the decrease in $E_{\rm Si-V}$. The bond potential does not account for interaction in the second coordination shell: having more silicon atoms around the vacancy is more energetically beneficial. Hence the $E_{\rm Si-V}$ increases for the bond potential. 

\begin{figure}
\centering
\begin{subfigure}[b]{0.45\linewidth}
    \includegraphics[width=\linewidth]{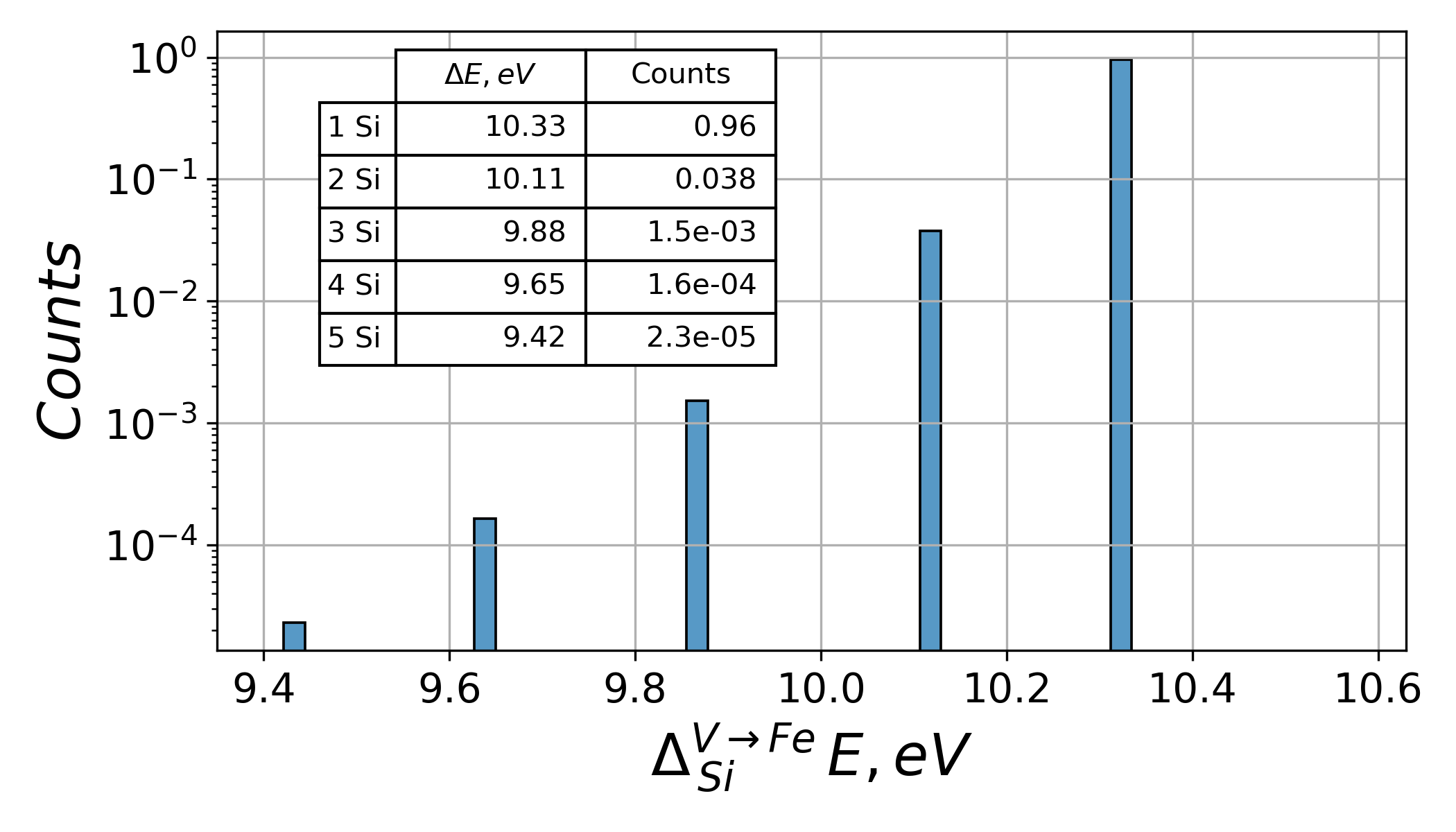}
    \caption{}
    \label{fig:EnerSpec_TL}
\end{subfigure} \quad
\begin{subfigure}[b]{0.45\linewidth}
    \includegraphics[width=\linewidth]{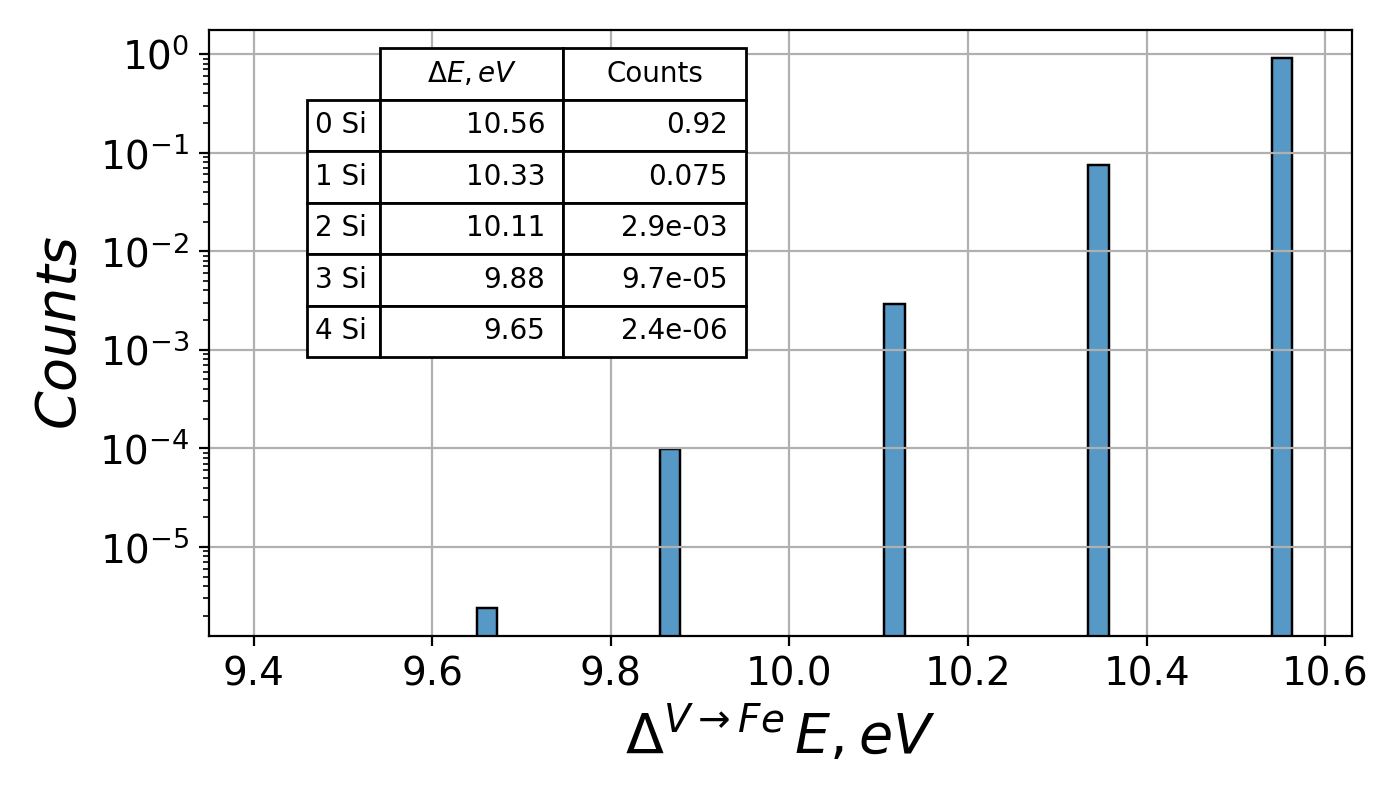}       
    \caption{}
    \label{fig:EnerSpec_TR}
\end{subfigure} \\
\begin{subfigure}[b]{0.45\linewidth}
    \centering
    \includegraphics[width=\linewidth]{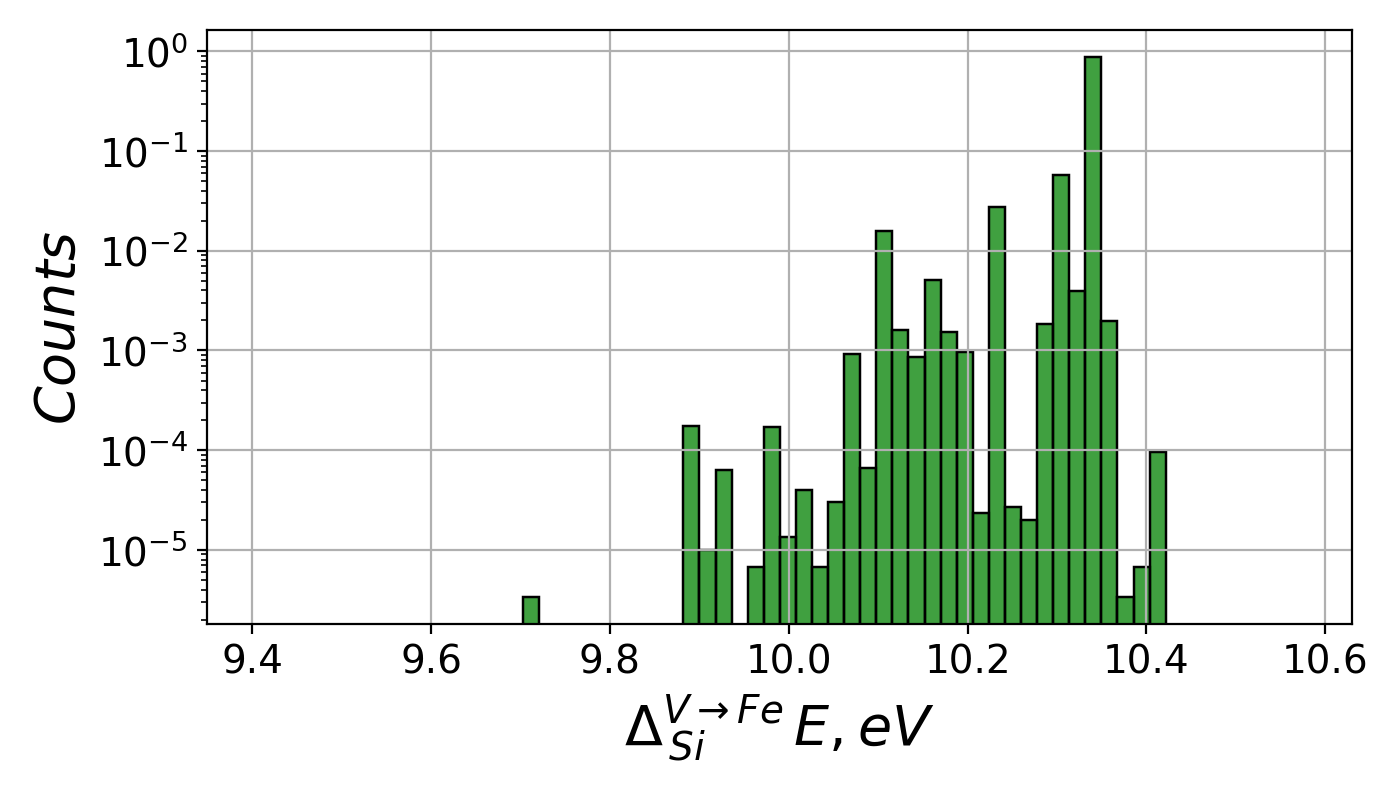}
    \caption{}
    \label{fig:EnerSpec_BL}
\end{subfigure}\quad
\begin{subfigure}[b]{0.45\linewidth}
    \centering
    \includegraphics[width=\linewidth]{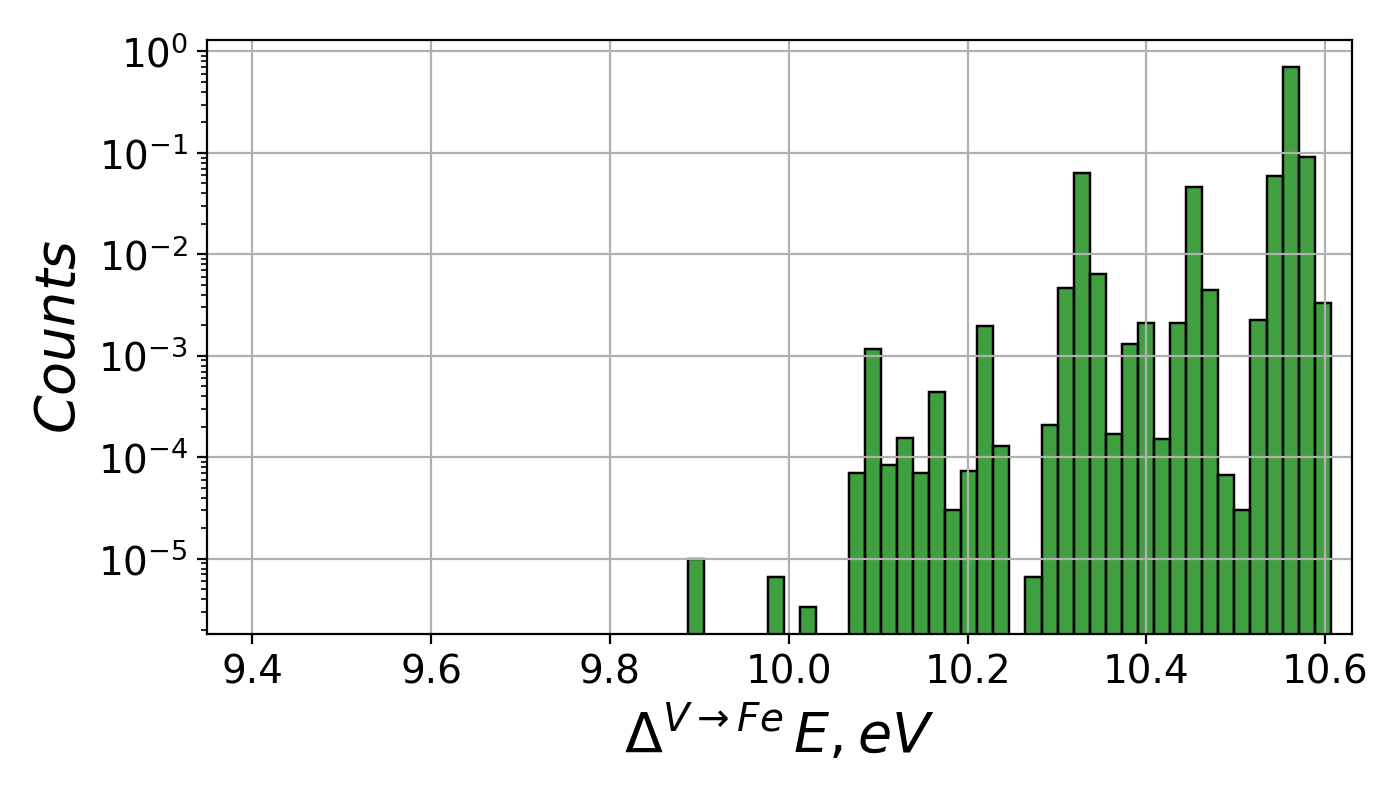}
    \caption{}
    \label{fig:EnerSpec_BR}
\end{subfigure}\quad
\caption{The substitution energy spectra of $ \Delta^{V \rightarrow Fe} E $ and $ \Delta^{V \rightarrow Fe}_{Si} E $: $\Delta^{V \rightarrow Fe} E$ - substitution energy of Fe to vacancy, $\Delta^{V \rightarrow Fe}_{Si} E$ is the substitution energy of Fe to vacancy, where Fe has solute atom among nearest neighbors. We calculate spectra using the bond potential (upper panels) and the low-rank potential (lower panels). Si concentration is equal to 1 at. \%.}
\label{fig:LRP_QChem_EnerSpectrum} 
\end{figure}

Figure \ref{fig:LRP_QChem_EnerSpectrum} shows the spectra of substitution energies: $\Delta^{\rm V \rightarrow Fe} E$ is the substitution energy of Fe to vacancy, $\Delta^{\rm Vac \rightarrow Fe}_{\rm Si} E$ is the substitution energy of Fe to vacancy, where Fe has solute atom among nearest neighbors. These substitution energies are further used to produce Si-Vac binding energy according to Equations \eqref{eq:eff_bind_energy}, \eqref{eq:mu_V_sub_ener}, \eqref{eq:mu_V-Si_sub_ener}. The bond potential spectra have sufficiently lower energy peaks compared with the low-rank potential spectra, as we expect from the previous analysis. These low energy peaks averaged according to Equation \eqref{eq:deltaE_Averaging} provide the dominant contribution to the $E_{\rm Si-V}$ values obtained with the bond potential.

The Si atoms do not interact with each other in the second coordination shell for the bond potential. After the vacancy exchanges with the Si atom, this Si atom is located in the first coordination shell of the previous vacancy neighbors. The state before the jump is energetically beneficial, but after the jump, Si atoms repel with significant energy (see Table \ref{tab:energiesBondModel}) making this state unbeneficial for the system. We account for this by introducing the linear correction \cite{Voter_KMC, kmc_mig_barrier_2} for the migration barrier: we take the difference between the final and initial energy states divided by half, see Equation \eqref{eq:delta_e_mig}.

The Fe-Si alloy has the body-centered cubic lattice with the eight sites in the first coordination shell; thus, there are eight unique energy configuration determined by the number of the silicon atoms ($N_{\rm Si}$) in the vacancy's first coordination shell. The probability of each energy configuration is unequal and decreases with the increase of $N_{\rm Si}$, as shown in the upper panels of Figure \ref{fig:LRP_QChem_EnerSpectrum}. We explicitly account for these probabilities $p(N_{\rm Si})$ in Equation \eqref{eq:nu_si_bond}. Therefore, the improved formula for the bond potential is the sum over the unique configurations with corresponding probabilities as the weight in the sum. Sum terms explicitly capture how the binding energy ($E_{\rm Si-V}(N_{\rm Si})$) and the migration energy correction ($\Delta E_{\rm mig}(N_{\rm Si})$) depend on the $N_{\rm Si}$.

Results by the improved formula are presented in Figure \ref{subfig:nuSi_bondModel_2}. We see that the results for the Si jump frequency agree with the kinetic simulations, so the improved Equation \eqref{eq:nu_si_bond} captures the physics of the process more accurately.

\newpage
\subsection{Correlation Factor}
\label{subsec:f2_si}

The correlation factor ($f_{Si}$) can not be estimated from thermodynamic simulations, as it is a purely kinetic property. Analytical five-frequency and nine-frequency models estimate $f_{Si}$ for the case $C_{Si} \rightarrow 0$. This section shows that the correlation factor only slightly depends on the silicon concentration. Thus, for approximate diffusion coefficient estimations, one can use analytical estimations.

\begin{figure}[h!]
\begin{center}
	\includegraphics[width=0.80\textwidth]{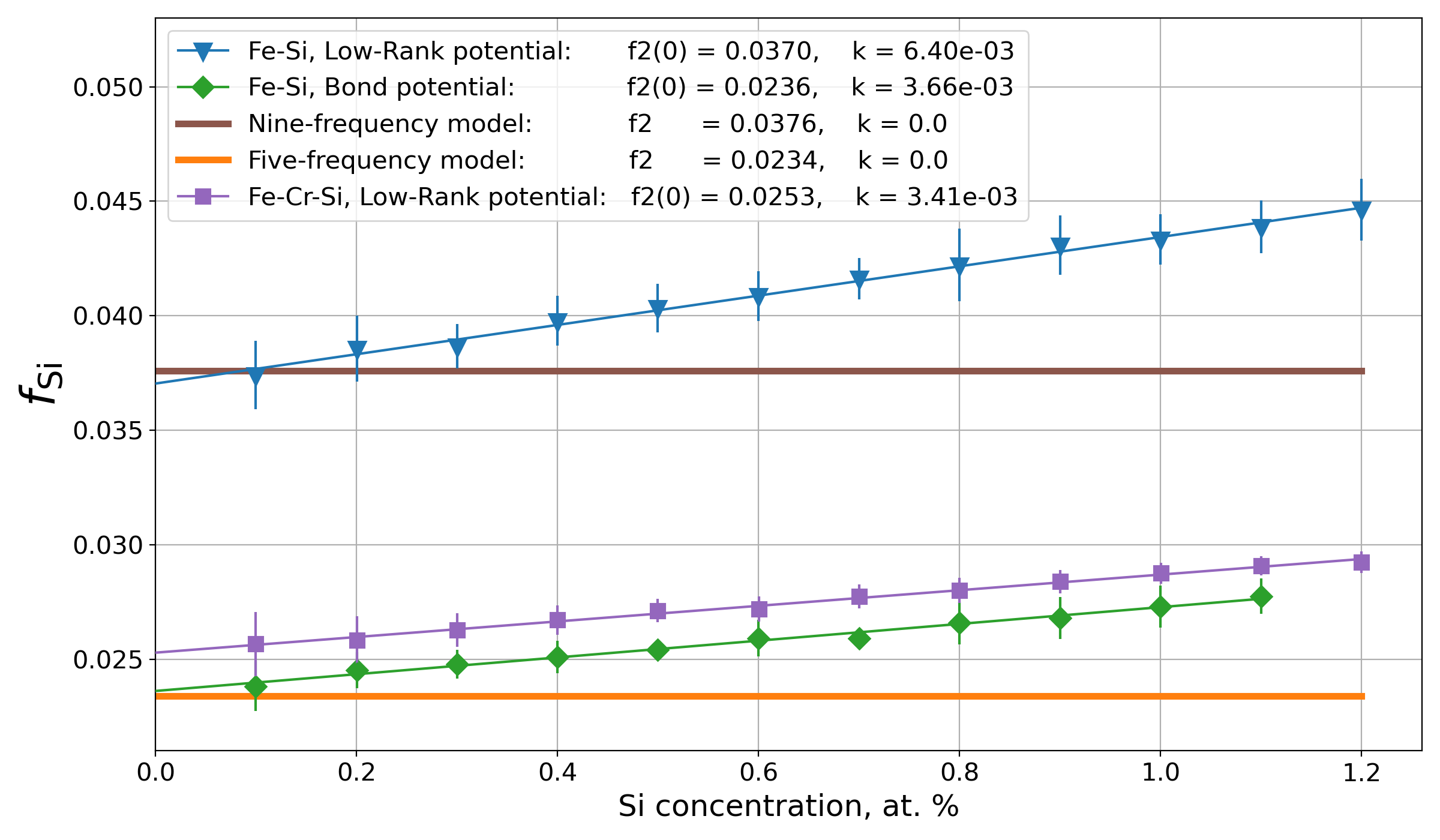}
	\caption{Silicon correlation factor estimated in kinetic Monte Carlo simulations with the low-rank potential and the bond potential. Horizontal lines denote the concentration-independent analytical nine-frequency and five-frequency models.}
	\label{fig:f2_FullPlot}
\end{center}
\end{figure}

We estimate the correlation coefficient from the kinetic Monte Carlo simulations. Figure \ref{fig:f2_FullPlot} shows how $ f_{Si} $ depends on Si concentration — increasing Si concentration from 0 to 1.2\% leads to 15-20 \% increase of $ f_{Si} $ in the Fe-Si alloy. Simulations with the bond potential in the Fe-Si system give two times lower values of $ f_{Si} $ compared to the reference calculations with the low-rank potential. This occurs because the low-rank potential accounts for more complex interactions than the bond potential. 

The nine-frequency \cite{Le_Claire_1970, LeClaire_Jones_1972} and five-frequency \cite{Manning1964} models are the analytical approaches that estimate solute correlation factor in the case of infinite dilution. We linearly extrapolate $f_{Si} (C_{Si}) $ for $C_{Si} \rightarrow 0$ to compare the simulation results with analytical models. The legend of Figure \ref{fig:f2_FullPlot} provides the extrapolation results, where $k$ is the curve slope. 

The nine-frequency model nicely agrees with the low-rank potential simulations, where the five-frequency model underestimates $f_{Si}$ by half. That is expected as the nine-frequency model is more sophisticated and accounts for vacancy jumps up to the fifth coordination shell. Five-frequency model considers interactions only between the nearest neighbors: thus, we see the agreement with the bond potential simulations as $C_{Si} \rightarrow 0$.   

In Figure \ref{fig:f2_FullPlot}, we also demonstrate the silicon correlation factor in Fe-Cr-Si alloy with Cr concentration 9 at. \%. The silicon correlation factor has a similar dependence as in the Fe-Si alloy: the correlation factor changes by nearly 15 \% when silicon concentration increases from 0 to 1.2 at \%. In the Appendix, we discuss stainless steel in detail. 

\newpage

\section{Conclusions}
\label{sec:conclusions}

We developed approaches to efficiently compute solute diffusion and test them on the Fe-Si alloy. In particular, we show how one can estimate solute jump frequency from thermodynamic modeling without performing dynamic simulations. To achieve this, we improve the formula for solute jump frequency to account for the local environment and migration energy variation for the bond potential. We demonstrate that the silicon correlation factor only slightly depends on silicon concentration; thus, the analytical nine-frequency model may be used without additional computations.

Also, we show that the computationally efficient bond potential may be used instead of the low-rank potential in the kinetic Monte Carlo simulations to estimate the silicon jump frequency. We demonstrate that the bond potential may be used in Fe-Si systems with Si concentrations up to 1 at. \% and in Fe-Cr-Si systems with Cr concentrations up to 10 at. \%.


\section{Appendix}

In our work, we focused on investigating the Fe-Si system. In industry, stainless steel has become widespread for its corrosion resistance properties. In stainless steel chromium is added to the Fe system, along with the small fractions of different chemical elements. Here, we investigate the Fe-Cr-Si system for Cr concentration up to 12 at. \%. 

\subsection{Appendix A: Silicon Jump Frequency in Stainless Steel}
\label{sec:nuSi_feCrSi}

\begin{figure}[h!]
\begin{center}
	\includegraphics[width=0.95\textwidth]{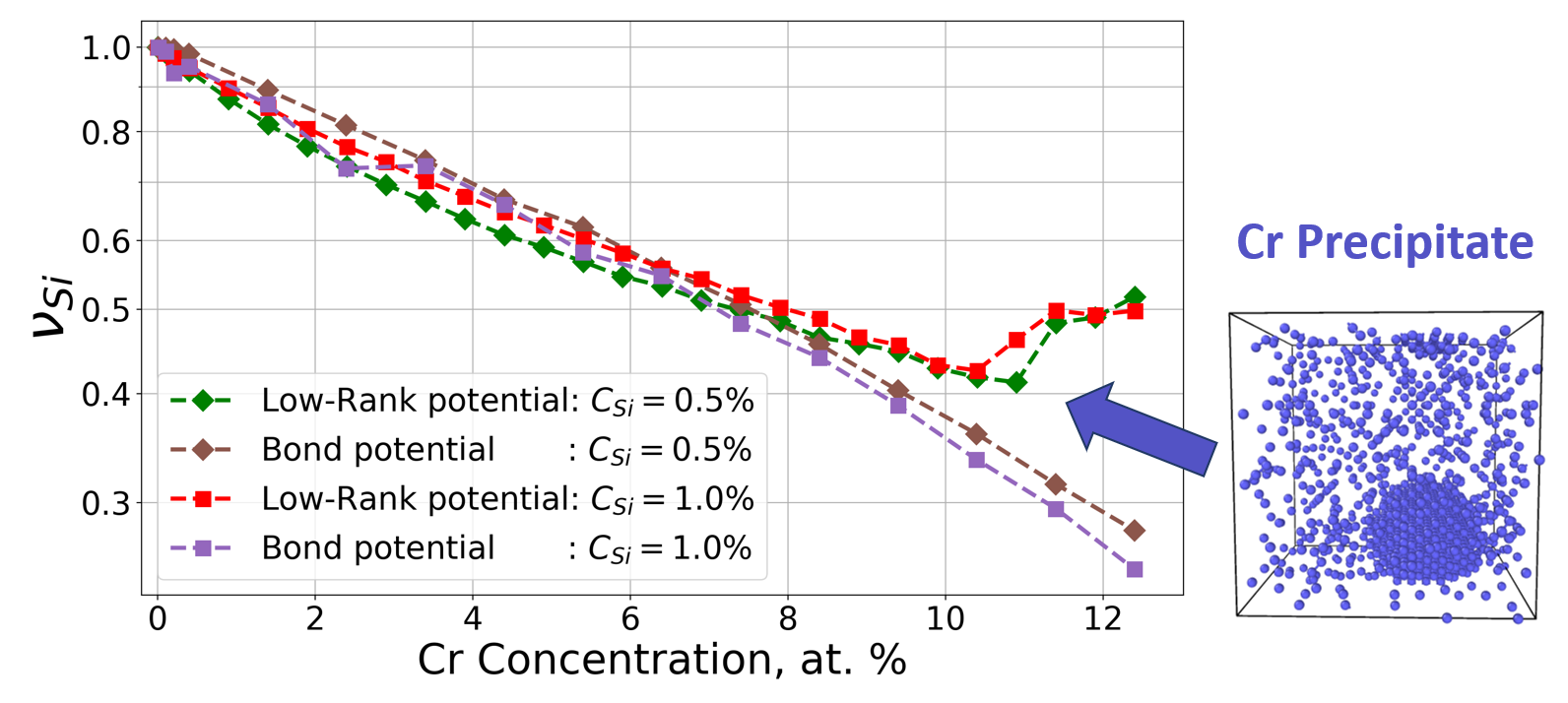}
	\caption{
	The silicon jump frequency obtained with the low-rank and bond potentials in the Fe-Cr-Si alloy. The silicon jump frequency ($\nu_{\rm Si}$) is normed to the frequency in the infinitely diluted system.
    }
	\label{fig:nuSi_3CompSys}   
\end{center}
\end{figure}

We show that the bond potential may be used instead of the low-rank potential to estimate silicon jump frequency in Fe-Cr-Si alloy. We plot the kinetic Monte Carlo simulations' results in Figure \ref{fig:nuSi_3CompSys} for silicon concentrations of 0.5 at. \% and 1.0 at. \%. The silicon jump frequency obtained with the bond potential is in good agreement with the low-rank potential up to 10 at. \% of Cr.

The silicon jump frequency makes the step change for the low-rank potential when the Cr concentration is higher than 10 at. \% because the Cr precipitates are formed. The experimental studies \cite{phase_diag_collection, FeCr_phaseTransition_exp_2008} report the precipitate formation for the Cr concentrations larger than 10 at. \%, which is in good agreement with our calculations. The bond potential is not designed to describe precipitates energies, which limits its application for the Cr concentrations lower than 10 at. \%.

\subsection{Appendix B: Correlation Factor in Stainless Steel}

\begin{figure}[h!]
    \begin{center}
        \includegraphics[width=0.8\textwidth]{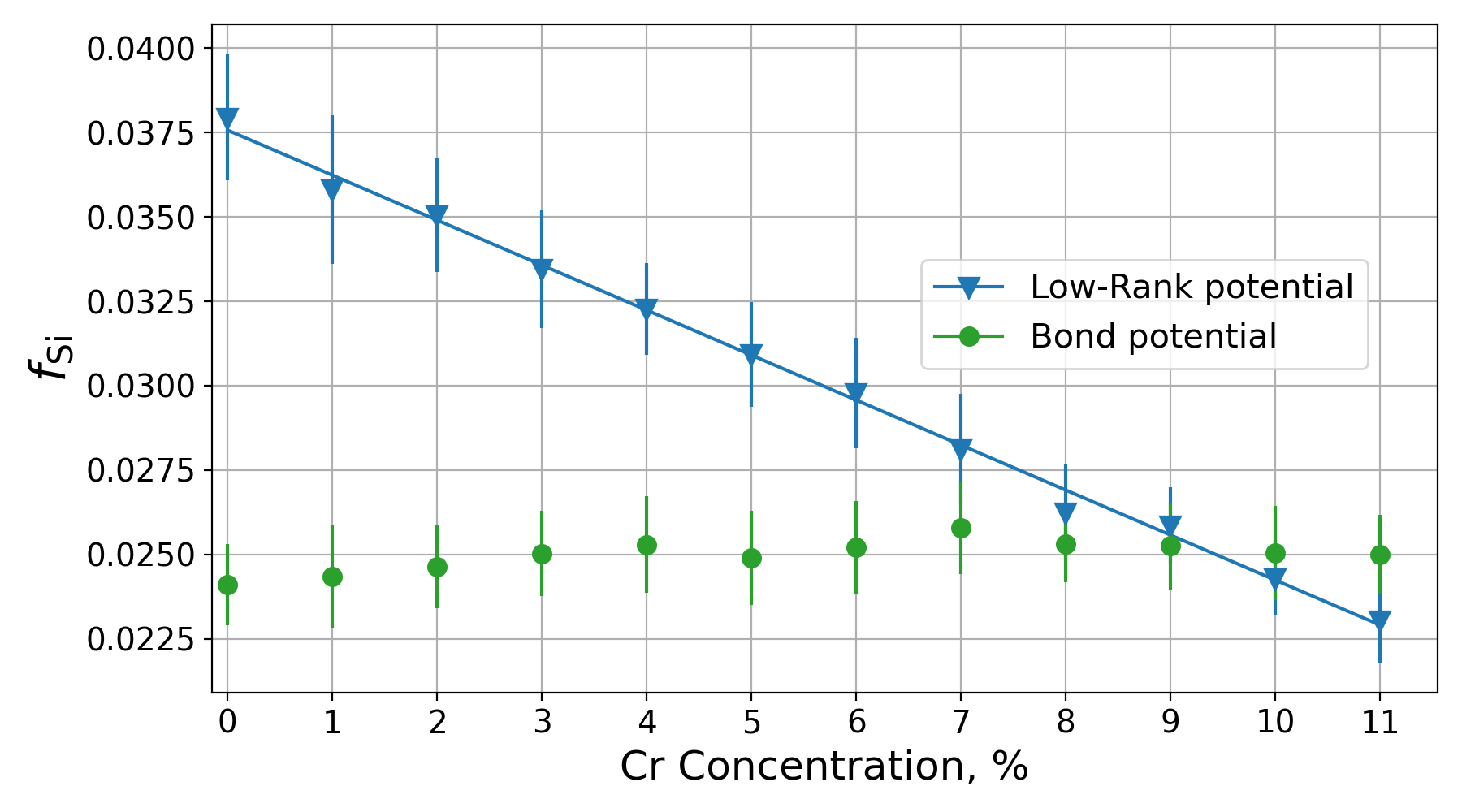}
        \caption{The silicon correlation factor in the Fe-Cr-Si alloy. The silicon concentration is equal to 0.1 at.\%.}
        \label{fig:f2Si_Cr_dep}
    \end{center}
\end{figure}

Figure \ref{fig:f2Si_Cr_dep} presents the silicon correlation factor in the Fe-Cr-Si alloy. The correlation factor linearly depends on the Cr concentration for the simulations with the low-rank potential. This dependence may be approximated by the linear function $f(C_{Si}) = b + k \cdot C_{Si} $ with the parameters $ b = 0.0376 $, $ k = -0.0013 $. The bond potential is not accurate enough to capture how the silicon correlation factor changes in the alloys with the sufficient chromium concentrations. We believe that this happens because the the atoms and vacancy interact in the second and third coordination shells (as shown in Figure \ref{fig:Eb_LRP_Ref_Comparas}), which the bond potential can not reproduce.

\newpage
\clearpage 
\bibliographystyle{IEEEtran}
\bibliography{biblio}

\end{document}